\documentclass[10pt]{article}

\usepackage{jheppub}
\usepackage{tikz-cd}
\usepackage{adjustbox}
\usepackage{pifont}
\usepackage{makecell}
\usepackage{orcidlink}

\usepackage{fontawesome5}
\makeatletter
\newcommand{\github}{{\href{https://github.com/HeChenWeng/Symbol-Alphabets-in-QCD-and-Flag-Cluster-Algebras.git}{\faGithub}}}
\makeatother

\newcommand{\la}{\langle}
\newcommand{\ra}{\rangle}
\newcommand{\Fl}{\mathcal{F\ell}}
\newcommand{\cmark}{\ding{51}\phantom{\ding{55}}}
\newcommand{\xmark}{\ding{55}\phantom{\ding{51}}}

\DeclareMathOperator{\Gr}{Gr}
\DeclareMathOperator{\sgn}{sgn}

\allowdisplaybreaks

\title{Symbol Alphabets in QCD and Flag Cluster Algebras}

\author[a, \orcidlink{0000-0003-1186-4624}]{Andrzej Pokraka,}
\emailAdd{andrzej\_pokraka@brown.edu}

\author[a,b, \orcidlink{0009-0005-6084-2466}]{Marcus Spradlin,}
\emailAdd{marcus\_spradlin@brown.edu}

\author[a, \orcidlink{0009-0008-2506-3207}]{Anastasia Volovich,}
\emailAdd{anastasia\_volovich@brown.edu}

\author[a, \orcidlink{0000-0003-3781-6153}]{and He-Chen Weng}
\emailAdd{he-chen\_weng@brown.edu}

\affiliation[a]{Department of Physics,
    Brown University,
    Providence,
    RI 02912,
    USA
}

\affiliation[b]{Brown Theoretical Physics Center,
    Brown University,
    Providence,
    RI 02912,
    USA
}

\abstract{The full 245-letter symbol alphabet for all planar massless two-loop six-point Feynman integrals was recently determined in \href{https://arxiv.org/abs/2412.19884}{arXiv:2412.19884} and \href{https://arxiv.org/abs/2501.01847}{arXiv:2501.01847}. In a parallel mathematical development, it was shown in \href{https://arxiv.org/abs/2408.14956}{arXiv:2408.14956} that there is an embedding of the cluster algebra associated to the partial flag variety $\Fl_{2,n-2;n}$, which describes the kinematics of $n$ massless particles, into that of the Grassmannian $\Gr(n{-}2,2n{-}4)$. In this paper we connect these developments by showing that most of the rational symbol letters can be expressed in terms of flag cluster variables, and that all of the algebraic symbol letters arise from infinite mutation sequences.}

\begin{document}

\maketitle

\section{Introduction}

A better understanding of the analytic properties of perturbative scattering amplitudes has enabled numerous breakthroughs in recent years, both conceptual and technological. In particular, it is now appreciated that there is deep mathematics underlying the patterns of singularities of amplitudes in kinematic space. One aspect of this is the connection between symbol letters~\cite{Goncharov:2010jf} of amplitudes in planar $\mathcal{N}=4$ super-Yang-Mills (SYM) theory and Grassmannian cluster algebras~\cite{fomin2002cluster,gekhtman2003cluster,scott2006grassmannians}, first observed for two-loop six- and seven-point amplitudes in~\cite{Golden:2013xva}. The hypothesis that this remains true to all loop order underlies a symbol bootstrap program~\cite{Dixon:2011pw,Caron-Huot:2016owq,Caron-Huot:2019vjl} (see~\cite{Caron-Huot:2020bkp} for a review) that has enabled the six-point MHV amplitude to be computed to eight loops~\cite{Dixon:2023kop}.

The fact that cluster algebras play a role in dictating the singularities of amplitudes is still somewhat mysterious, but the fact that the cluster algebras relevant to SYM theory are of Grassmannian type is natural in light of the momentum twistor parameterization~\cite{Hodges:2009hk} of its kinematic space in terms of collections of ordered points in $\mathbb{P}^3$. The question of whether other cluster algebras might be relevant to the singularities of amplitudes or integrals in other quantum field theories, with more complicated kinematic spaces, was answered in the affirmative in~\cite{Chicherin:2020umh}, which found a kinematic parameterization in which the symbol letters of all two-loop four-point master integrals with one off-shell (or equivalently massive) leg~\cite{Gehrmann:2000zt,Gehrmann:2001ck} are cluster variables of the $C_2$ cluster algebra. These master integrals enter the calculation of Higgs plus jet production in the heavy-top limit of QCD. Additional three-loop planar integral families with the same structure were found in~\cite{Canko:2021hvh,Canko:2021xmn}, and it was shown in~\cite{Aliaj:2024zgp} that the singularities of a non-planar three-loop integral evaluated in~\cite{Henn:2023vbd} are described (again, after a suitable parameterization of the relevant kinematic space) by the $G_2$ cluster algebra. Note that $C_2$ is a subalgebra of $G_2$ (and the relevant one-loop integrals are described by $A_2$, which in turn is a subalgebra of $C_2$), but neither $C_2$ nor $G_2$ are of Grassmmanian type.

The perturbative amplitude bootstrap has seen very fruitful application also outside of SYM theory. In particular, the 26-letter symbol alphabet for all planar massless two-loop five-point master integrals was introduced in~\cite{Gehrmann:2015bfy} (the corresponding function space was constructed in~\cite{Gehrmann:2018yef}), and extended to a 31-letter alphabet for the non-planar sector in~\cite{Chicherin:2017dob}. These results have enabled tremendous progress in analytical calculations of integrals and amplitudes of direct relevance for QCD~\cite{Chicherin:2018mue,Chicherin:2018old,Chicherin:2020oor} as well as SYM theory~\cite{Abreu:2018aqd} and $\mathcal{N}=8$ supergravity~\cite{Abreu:2019rpt}. The complete alphabet and function space for planar six-point integrals at two loops has very recently been determined in~\cite{Abreu:2024fei,Henn:2025xrc}.

In light of these developments, it is always interesting to continue to explore the connection of cluster algebras to symbol alphabets of integrals beyond the special case of SYM theory.  The kinematic space parameterized by the spinor helicity variables associated to a scattering configuration of $n$ massless particles is isomorphic to the partial flag variety $\Fl_{2,n-2;n}$~\cite{Maazouz:2024qmm}. It is known that there are cluster algebra structures associated to partial flag varieties~\cite{Geiss2006PartialFV}, and it was recently shown~\cite{Bossinger:2024apm} that there is an embedding of these into cluster structures associated to Grasmannians.  In the case of relevance to $n$-particle massless amplitudes, their construction shows that the initial cluster of $\Fl_{2,n-2;n}$ can be obtained from that of $\Gr(n{-}2,2n{-}4)$ by a sequence of $\frac{(n-5)(n^2-10 n+30)}{2}$ mutations followed by freezing $n{-}5$ variables and deleting $(n{-}3)(n{-}5)$ nodes.

In this paper we explore the connection between the $\Fl_{2,n-2;n}$ cluster algebra and massless two-loop $n$-point symbol alphabets for $n=5, 6$ (restricted to the planar case for $n=6$, but including non-planar integrals for $n=5$). For $n=5$, 30 of the 31 symbol letters correspond precisely to (permutations of) cluster variables of $\Fl_{2,n-2;n}$, and the remaining letter $W_{31}$ appears in individual Feynman integrals but (as far as is currently known) drops out of every suitably defined finite quantity in four dimensions. This observation was already made in an essentially identical form by Bossinger, Drummond and Glew in~\cite{Bossinger:2022eiy}; there it was phrased in terms of Gr\"obner fans, though the connection to partial flag varieties was also explained in~\cite{2638233}.  For $n=6$ several new features emerge; most notably, the cluster algebra $\Fl_{2,4;6}$ is infinite, and one can use the construction of~\cite{Drummond:2019cxm,ccanakcci2018cluster} to naturally associate certain algebraic letters to the algebra, one example of which was already presented in~\cite{2638233}.

The structure of this paper is as follows. In Section~2 we review how the $\Fl_{2,3;5}$ cluster algebra encodes all but one of the 31 letters of the two-loop five-point symbol alphabet of~\cite{Chicherin:2017dob}. In Section~3 we review the 245-letter planar two-loop six-point symbol alphabet of~\cite{Abreu:2024fei,Henn:2025xrc} and show that all 40 of its algebraic letters, and 135 of its rational letters, are associated to the $\Fl_{2,4;6}$ cluster algebra.  In Section~4 we summarize our results and indicate several directions for future work.

\bigskip
\noindent
{\bf Note added.} While we were finishing this paper we became aware of partially overlapping results obtained by L.~Bossinger, J.~Drummond, R.~Glew, \"O.~G\"urdo\u{g}an and R.~Wright~\cite{Bossinger:2025rhf,RWrightCMSA}.

\section{Two-loop Five-point Integrals}

\subsection{Review of the symbol alphabet}
\label{sec:fivePtalphabet}

The kinematic space for massless five-point scattering consists of the Lorentz invariants formed from five momentum vectors $p_1,\ldots,p_5$ subject to the on-shell conditions $p_i^2 = 0$ and momentum conservation $\sum_{i=1}^5 p_i = 0$. There are ten distinct nonzero scalar products. We choose a basis given by the five variables
\begin{align}\begin{aligned}
v_1 &= (p_1 + p_2)^2\,, &v_{i\in[2,5]} = T^{i-1}v_1\,, \\
\end{aligned}\end{align}
where $T$ is the generator of cyclic permutations: $T p_i = p_{(i+1)\,\text{mod}\,5}$. In addition to the $v_i$, which are parity-even, there is also a parity-odd Lorentz invariant that is naturally expressed in terms of
spinor helicity variables as
\begin{align}
\epsilon_{1234} = \langle12\rangle\langle45\rangle[24][15]-\langle24\rangle\langle15\rangle[12][45]\,.\label{eq:fivepointodd}
\end{align}

Here we review a basis for the complete set of 31 multiplicatively independent letters\footnote{Of course, symbol letters must be dimensionless so one can always choose a basis of at most 30 multiplicatively independent letters.} that appear in massless two-loop five-point Feynman integrals, to arbitrary order in dimensional regularization~\cite{Gehrmann:2015bfy,Chicherin:2017dob}.  First we have 25 letters that are linear in Mandelstam variables:
\begin{align}\begin{aligned}
    W_1 &= v_1
    \,,
    &
    W_{i\in[2,5]} &= T^{i-1} W_{1}
    \,,
    &
    W_6 &= v_3{+}v_4
    \,,
    &
    W_{i\in[7,10]} &= T^{i-6} W_{6}
    \,,
    \\
    W_{11} &= v_1{-}v_4
    \,,
    &
    W_{i\in[12,15]} &= T^{i-11} W_{11}
    \,,
    &
    W_{16} &= v_4{-}v_1{-}v_2
    \,,
    &
    W_{i\in[17,20]} &= T^{i-16} W_{16}
    \,,
    \\
    W_{21}&=v_3{+}v_4{-}v_1{-}v_2
    \,
    & W_{i\in[22,25]} &= T^{i-21} W_{21}\,.
\end{aligned}\end{align}
The remaining 6 letters are not rational functions of Mandelstam variables, but are rational when expressed in terms of spinor helicity variables:
\begin{align}\begin{aligned}
    W_{26}&=\frac{v_1v_2{-}v_2v_3{+}v_3v_4{-}v_4v_5{-}v_5v_1{-}\epsilon_{1234}}{v_1v_2{-}v_2v_3{+}v_3v_4{-}v_4v_5{-}v_5v_1{+}\epsilon_{1234}}
    \,,
    &
    W_{i\in[27,30]} &= T^{i-26} W_{26}
    \,,
    \\
    W_{31} &= \epsilon_{1234}
    \,.
\end{aligned}\end{align}
If one restricts to functions that result from planar Feynman integrals, the corresponding alphabet does not include the letters $W_{[21,25]}$.

\subsection{\texorpdfstring{The $\Fl_{2,3;5} \cong D_4$ cluster algebra}{The Fl2,3;5 = D4 cluster algebra}}
\label{sec:fl235}

In this section, we review how to construct the five-point alphabet from the cluster algebra associated to the partial flag variety $\mathcal{F\ell}_{2,3;5}$.
This construction was first worked out in~\cite{Bossinger:2022eiy} from the perspective of Gr\"obner fans, and the connection to the partial flag variety was also noted in the thesis~\cite{2638233}.

A point in the partial flag variety $\mathcal{F\ell}_{2,3;5}$ is a pair of vector spaces $V_2\subset V_3\subset\mathbb{C}^5$, with $\dim V_d = d$, which can be represented by (respectively) the row span of the first 2 and 3 rows of a $3\times 5$ matrix $M$. The Pl\"ucker coordinates on $\mathcal{F\ell}_{2,3;5}$ are the $2\times2$ and $3\times3$ minors of $M$:
\begin{equation}
    P_{ i,j} = \det\left(M_{a,b}\right)_{a \in \{1,2\},\ b\in \{i,j\}}\text{ and }P_{i,j,k} := \det\left(M_{a,b}\right)_{a \in \{1,2,3\},\ b\in \{i,j,k\}}.
\end{equation}
The explicit mapping between the spinor helicity brackets of massless five-point kinematics and the the Pl\"ucker coordinates of $\mathcal{F\ell}_{2,3;5}$ is simply
\begin{equation}
\label{eq:fivepointcorrespondence}
    \langle ij \rangle \leftrightarrow P_{i,j}\quad \text{and}\quad [ij]\leftrightarrow (-1)^{i+j+1}P_{\{1,\dots,5\}\setminus \{i,j\}}
\end{equation}
for $i<j$.

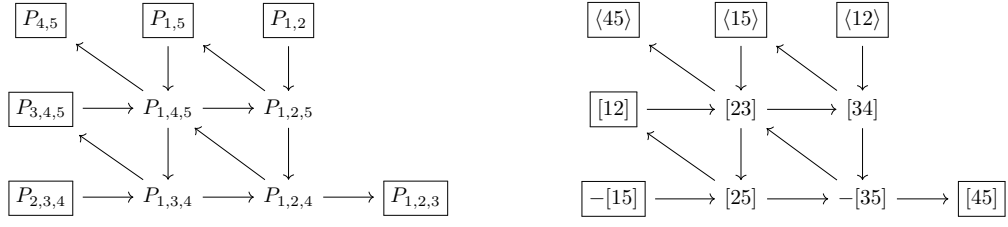
\begin{figure}
     \centering
\adjustbox{scale=.8}{\begin{tikzcd}
 	\boxed{P_{4,5}} & {\boxed{P_{1,5}}} & {\boxed{P_{1,2}}} \\
 	\boxed{P_{3,4,5}} & {P_{1,4,5}} & {P_{1,2,5}} \\
 	\boxed{P_{2,3,4}} & {P_{1,3,4}} & {P_{1,2,4}} & {\boxed{P_{1,2,3}}}\\
        \arrow[from=2-1, to=2-2]
        \arrow[from=2-2, to=2-3]
        \arrow[from=3-1, to=3-2]
        \arrow[from=3-2, to=3-3]
        \arrow[from=1-2, to=2-2]
        \arrow[from=2-2, to=3-2]
        \arrow[from=1-3, to=2-3]
        \arrow[from=2-3, to=3-3]
        \arrow[from=2-2, to=1-1]
        \arrow[from=2-3, to=1-2]
        \arrow[from=3-2, to=2-1]
        \arrow[from=3-3, to=2-2]
        \arrow[from=3-3, to=3-4]
\end{tikzcd}}
\qquad\qquad
\adjustbox{scale=0.8}{\begin{tikzcd}
 	\boxed{\la 45 \ra} & {\boxed{\la15\ra}} & {\boxed{\la12\ra}} \\
 	\boxed{[12]} & {[23]} & {[34]} \\
 	\boxed{-[15]} & {[25]} & {-[35]} & {\boxed{[45]}}\\
        \arrow[from=2-1, to=2-2]
        \arrow[from=2-2, to=2-3]
        \arrow[from=3-1, to=3-2]
        \arrow[from=3-2, to=3-3]
        \arrow[from=1-2, to=2-2]
        \arrow[from=2-2, to=3-2]
        \arrow[from=1-3, to=2-3]
        \arrow[from=2-3, to=3-3]
        \arrow[from=2-2, to=1-1]
        \arrow[from=2-3, to=1-2]
        \arrow[from=3-2, to=2-1]
        \arrow[from=3-3, to=2-2]
        \arrow[from=3-3, to=3-4]
\end{tikzcd}}
\caption{The initial cluster of $\mathcal{F\ell}_{2,3;5}$ with nodes labeled by the Pl\"ucker coordinates of the partial flag variety (left) or by the corresponding spinor helicity variables (right), with frozen nodes boxed.}
     \label{figure: initial cluster 235}
\end{figure}

The prescription for writing down an initial cluster for partial flag variety cluster algebras is given in~\cite{Bossinger:2024apm}. For the special case of $\Fl_{2,3;5}$ relevant to massless five-point scattering, the initial cluster, shown in Figure~\ref{figure: initial cluster 235}, happens to be isomorphic to that of the Grassmannian $\Gr(3,6)$, whose cluster algebra is of type $D_4$. The correspondence between the Pl\"ucker coordinates $P_{i,j}$, $P_{i,j,k}$ of the partial flag variety and those of the Grassmannian, which we denote $\langle ijk \rangle$, is
\begin{align}
\label{eq:cyclicdecoupling}
P_{i,j} \leftrightarrow \langle ij6 \rangle \qquad
P_{i,j,k} \leftrightarrow \langle ijk \rangle\,.
\end{align}

The set of all clusters is obtained by applying all possible mutation sequences to the initial cluster; see~\cite{Fomin:2016caz} for an elementary introduction to cluster algebras and, in particular, the mutation rules.
Altogether there are 50 distinct clusters, containing a total of 22 distinct cluster variables.
This set includes all 10 two-entry Pl\"ucker coordinates and all 10 three-entry Pl\"ucker coordinates, which correspond via~(\ref{eq:fivepointcorrespondence}) to all spinor brackets $\langle i j \rangle$ and $\pm[ij]$.
(Since our aim is to describe symbol letters, overall signs are immaterial.)
Additionally, there are two cluster variables quadratic in Pl\"ucker coordinates, which correspond in spinor helicity variables to
\begin{align}\begin{aligned}
    P_{2,3}P_{1,4,5}-P_{4,5}P_{1,2,3}
    &\leftrightarrow
    \la 23\ra [23]{-}\la 45\ra[45]
    \,,
    \\
    P_{3,4}P_{1,2,5}-P_{1,2}P_{3,4,5}
    &\leftrightarrow
    \la34\ra [34]{-}\la12\ra [12]
    \,.
\end{aligned}\end{align}

\subsection{\texorpdfstring{Symbol letters from $\Fl_{2,3;5}$ cluster variables}{Symbol letters from Fl2,3;5 cluster variables}}

This set of 22 cluster variables associated to $\Fl_{2,3;5}$ is not closed under cyclic permutations $i \to (i{+}1)\,{\text{mod}}\,5$ of the five-point kinematic data, under which the symbol alphabet reviewed in Section~\ref{sec:fivePtalphabet} is of course closed. The origin of this mismatch is the correspondence~(\ref{eq:cyclicdecoupling}) which clearly decouples the $\mathbb{Z}_5$ cyclic symmetry of the kinematic space from the $\mathbb{Z}_6$ symmetry of $\Gr(3,6)$.

Motivated by these observations, we follow~\cite{Bossinger:2022eiy,2638233} in taking the closure of the set of
cluster variables under cyclic permutations of the indices, which provides the additional three letters needed to complete~(\ref{eq:cyclicdecoupling}) into a cyclic family. Moreover, in order to describe non-planar integrals, the symbol alphabet must be closed under the full $S_5$ permutation group.  Taking the closure of~(\ref{eq:cyclicdecoupling}) under arbitrary permutations produces another 10 letters that fall into two cyclic classes of five each.  In this manner, we obtain a total of 35 variables in seven cyclic families:
\begin{align}\begin{aligned}
    a_{1} &= P_{1,2}
    \leftrightarrow \la12\ra
    \,,
    &
    a_{i\in[2,5]}&=T^{i-1} a_1
    \,,
    \\
    a_{6} &= P_{1,3}
    \leftrightarrow \la13\ra
    \,,
    &
    a_{i\in[7,10]}&=T^{i-6} a_6
    \,,
    \\
    b_{1} &= P_{3,4,5}
    \leftrightarrow [12]
    \,,
    &
    b_{i\in[2,5]}&=T^{i-1} b_1
    \,,
    \\
    b_{6} &= P_{2,4,5}
    \leftrightarrow -[13]
    \,,
    &
    b_{i\in[7,10]}&=T^{i-6} b_6
    \,,
    \\
    c_{1} &= {-}P_{3,5} P_{1,2,4}{-}P_{1,2} P_{3,4,5}
    \leftrightarrow \la35\ra [35] {-} \la12\ra [12]
    \,,
    &
    c_{i\in[2,5]} &= T^{i-1} c_{1}
    \,,
    \\
    c_6 &=P_{4,5} P_{1,2,3}-P_{1,2} P_{3,4,5}
    \leftrightarrow \la45\ra [45]{-}\la12\ra [12]
    \,,
    &
    c_{i\in[7,10]} &=T^{i-6} c_6
    \,,
    \\
    c_{11}&=P_{2,5} P_{1,3,4}-P_{1,4} P_{2,3,5}
    \leftrightarrow \la25\ra [25]{-}\la14\ra [14]
    \,,
    &
    c_{i\in[12,15]}&=T^{i-11} c_{11}
    \,,
\end{aligned}\end{align}
where, as before, $T$ generates cyclic permutations of the indices of the Pl\"ucker coordinates according to $i\rightarrow (i{+}1)\,\text{mod}\,5$.

Expressing the five-point alphabet reviewed in Section~\ref{sec:fivePtalphabet} in terms of spinor helicity variables then reveals that 30 of the 31 symbol letters are expressible as products of (permutations) of $\Fl_{2,3;5}$ cluster variables. Explicitly:
\begin{align}\begin{aligned}
    W_{1} &= \langle12\rangle[12] \leftrightarrow a_1~b_1\,,
    &W_{i\in[2,5]}&=T^{i-1}W_{1}\,,\\
    W_{6} &= \langle34\rangle[34]{+}\langle45\rangle[45] \leftrightarrow -c_{1}\,, &W_{i\in[7,10]}&=T^{i-6}W_{6}\,,\\
    W_{11} &= \langle12\rangle[12]{-}\langle45\rangle[45] \leftrightarrow -c_6\,, &W_{i\in[12,15]}&=T^{i-11} W_{11}\,,\\
    W_{16} &= \langle45\rangle[45]{-}\langle12\rangle[12]-\langle23\rangle[23]=\langle13\rangle[13] \leftrightarrow -a_6~b_6\,,\ &W_{i\in[17,20]}&= T^{i-16}W_{16}\,,\\
    W_{21}&=\langle34\rangle[34]{+}\langle45\rangle[45]{-}\langle12\rangle[12]{-}\langle23\rangle[23] = \langle25\rangle[25]{-}\langle14\rangle[14]\leftrightarrow c_{11}\,,
    & W_{i\in[22,25]} &= T^{i-21}W_{21}\,,\\
    W_{26} &= \frac{\langle12\rangle\langle45\rangle[24][15]}{\langle24\rangle\langle15\rangle[12][45]} \leftrightarrow -\frac{a_1~ a_4~ b_7~ b_5}{a_7~ a_5~ b_1~ b_4}\,,\ &W_{i\in[27,30]}&= T^{i-26}W_{26}\,.
\end{aligned}\label{eq:fivepointmap}
\end{align}
Here, we restrict to the kinematic region where $\epsilon_{1234} < 0$;
when $\epsilon_{1234}>0$, $W_{26}$ and its cyclic images map to the reciprocal of the form given here.

In summary, we have reviewed that taking the closure of the 22 cluster variables associated to $\Fl_{2,3;5}$ under the $S_5$ permutation group produces a total of 35 variables: the 20 spinor helicity brackets $\langle ij \rangle$, $[ij]$ and 15 quadratic polynomials in brackets.  From this collection one can form a total of 30 multiplicatively independent cross ratios that are invariant under little group scaling of the spinor helicity variables.  Modulo overall signs, the multiplicative span of these 30 variables is almost the same as that of the five-point alphabet; the only symbol letter excluded is $W_{31} = \epsilon_{1234}$.  This is notable because although $W_{31}$ appears in individual Feynman integrals (starting at $\mathcal{O}(\epsilon^0)$ in dimensional regularization), it is not known to appear in any suitably defined finite amplitude; in particular it drops out from the hard part of the two-loop five-point amplitudes in $\mathcal{N} = 4$ SYM theory and $\mathcal{N}=8$ supergravity. An optimist might consider this amusing fact to be circumstantial evidence in favor of the ``correctness'' of this approach.

\section{Planar Two-loop Six-point Integrals}

\subsection{Review of the symbol alphabet}
\label{sec:sixpointalphabet}

The kinematic space for massless six-point scattering consists of the Lorentz invariants formed from six momentum vectors $p_1,\ldots,p_6$ subject to the on-shell conditions $p_i^2 = 0$ and momentum conservation $\sum_{i=1}^6 p_i = 0$. There are twenty-five distinct nonzero scalar products: fifteen  $(p_i+p_j)^2$ and ten $(p_i+p_j+p_k)^2$. We choose a basis given by the nine variables
\begin{align}\begin{aligned}
v_1 &= (p_1 + p_2)^2\,, &v_{i \in [2,6]} = T^{i{-}1}v_1\,, \\
v_7 &= (p_1 + p_2 + p_3)^2\,, &v_{i \in [8,9]} = T^{i{-}7}v_7\,,
\end{aligned}\end{align}
where $T$ is the generator of cyclic permutations: $T p_i = p_{(i+1)\,\text{mod}\,6}$. In addition to the $v_i$, which are parity-even, there are also parity-odd Lorentz invariants generalizing~(\ref{eq:fivepointodd}) that are naturally expressed in terms of spinor helicity variables as
\begin{align}\begin{aligned}
    \epsilon_{ijkl} &=
   [ij] \la jk\ra [kl] \la li\ra
    - \la ij\ra [jk] \la kl\ra [li]
    \,,\\
    \Delta_6 &= \la 12 \ra [23] \la34\ra [45] \la56\ra [61]
    - [12] \la23\ra [34] \la45\ra [56] \la61\ra\,.
\end{aligned}\end{align}

Here we review a basis for the complete set of 245 (or 244; see footnote~1) multiplicatively independent letters that appear in planar massless two-loop six-point Feynman integrals, to arbitrary order in dimensional regularization~\cite{Abreu:2024fei, Henn:2025xrc}; our numbering of the letters $W_i$ follows the latter reference. Note that~\cite{Henn:2025xrc} uses a multiplicatively overcomplete set of letters $W_{[1,289]}$; the basis we enumerate here omits the redundant letters $W_{[191,193] \cup [212,217] \cup [221,223] \cup [248,274] \cup [279,280] \cup [287,289]}$.

The 48 letters linear in Mandelstam variables are:
\begin{align}\begin{aligned}
    W_1
    &= v_1
    \,,
    &
    W_{i \in [2,6]} &= T^{i{-}1} W_1
    \,,
    \\
    W_7
    &= v_7
    \,,
    & W_{i\in [8,9]} &= T^{i{-}7} W_7
    \,,
    \\
    W_{10}
    &= {-}v_1{-}v_2
    \,,
    & W_{i\in [11,15]} &= T^{i{-}10} W_{10}
    \,,
    \\
    W_{16}
    &= v_1{-}v_7
    \,,
    &
    W_{i\in[17,21]} &= T^{i{-}16} W_{16}
    \,,
    \\
    W_{22}
    &= v_1{-}v_9
    \,,
    &
    W_{i\in[23,27]} &= T^{i{-}22} W_{22}
    \,,
    \\
    W_{28}
    &= v_7{-}v_1{-}v_2
    \,,
    &
    W_{i\in[29,33]} &= T^{i{-}28} W_{28}
    \,,
    \\
    W_{34}
    &= v_1{-}v_3{-}v_7
    \,,
    & W_{i\in [35,39]} &= T^{i{-}34} W_{34}
    \,,
    \\
    W_{40}
    &= v_1{-}v_5{-}v_9
    \,,
    & W_{i\in [41,45]} &= T^{i{-}40} W_{40}
    \,,
    \\
    W_{46}
    &= v_1{+}v_4{-}v_7{-}v_9
    \,,
    & W_{i\in [47,48]} &= T^{i{-}46} W_{46}
    \,.
\end{aligned}\label{eq:type1}\end{align}

The 51 letters quadratic in Mandelstam variables are:
\begin{align}
\begin{aligned}
    W_{49}
    &= v_7 v_9-v_1 v_4
    \,,
    & W_{i\in [50,51]} &= T^{i{-}49} W_{49}
    \,,
    \\
    W_{52}
    &= v_1 v_5{-}v_1 v_7{+}v_3 v_7
    \,,
    & W_{i\in [53,57]} &= T^{i{-}52} W_{52}
    \,,
    \\
    W_{58}
    &= v_7 v_9 {-}v_1 v_4{-}v_2 v_9
    \,,
    & W_{i\in [59,63]} &= T^{i{-}58} W_{58}
    \,,
    \\
    W_{64}
    &= v_7 v_9{-}v_1 v_4{-}v_3 v_7
    \,,
    & W_{i\in [65,69]} &= T^{i{-}64} W_{64}
    \,,
    \\
    W_{70}
    &= v_1 v_5{-}v_7 v_5{+}v_3 v_7
    \,,
    & W_{i\in [71,75]} &= T^{i{{-}}70} W_{70}
    \,,
    \\
    W_{76}
    &= v_1 v_3{-}v_9 v_3{+}v_1 v_4{+}v_5 v_9{-}v_7 v_9
    \,,
    & W_{i\in [77,81]} &= T^{i{{-}}76} W_{76}
    \,,
    \\
    W_{82}
    &= v_1 v_4{+}v_2 v_4{-}v_2 v_7{+}v_6 v_7{-}v_7 v_9
    \,,
    & W_{i\in [83,87]} &= T^{i{{-}}82} W_{82}
    \,,
    \\
    W_{88}
    &= {-}v_9^2{+}v_1 v_9{+}v_3 v_9{-}v_5 v_9{-}v_1 v_3
    \,,
    & W_{i\in [89,93]} &= T^{i{{-}}88} W_{88}
    \,,
    \\
    W_{94}
    &= {-}v_1^2{+}2 v_3 v_1{+}v_7 v_1{-}v_8 v_1{-}v_3^2{-}v_2 v_5{-}v_3 v_7{+}v_3 v_8{+}v_7 v_8
    \,,
    & W_{i\in [95,99]} &= T^{i{{-}}94} W_{94}
    \,.
\end{aligned}\label{eq:type2}
\end{align}
The 18 letters cubic in Mandelstam variables are:
\begin{align}
\begin{aligned}
    W_{100}
    &= v_2v_5(v_9{-}v_3)
     {+}v_1v_3(v_8{-}v_6)
     {+}v_7v_8(v_3{-}v_9)
     {+}v_6v_7(v_9{-}v_3)
    \,,
    &
    W_{i\in [101,105]} &= T^{i{-}100} W_{100}
    \,,
    \\
    W_{106}
    &= v_7 v_5^2{-}v_1 v_2 v_5{-}v_3 v_7 v_5{+}v_1 v_8 v_5{-}v_7 v_8 v_5{+}v_3 v_7 v_8{-}v_1 v_5^2
    \,,
    & W_{i\in [107,111]} &= T^{i{-}106} W_{106}
    \,,
    \\
    W_{112}
    &= v_8 v_2^2{-}v_5 v_6 v_2{+}v_6 v_7 v_2{-}v_4 v_8 v_2{-}v_7 v_8 v_2{-}v_2^2 v_6{+}v_4 v_7 v_8
    \,,
    & W_{i\in [113,117]} &= T^{112-i} W_{112}
    \,.
\end{aligned}\label{eq:type3}
\end{align}
(Note the inversion of $T$ in the last line.)

The remaining letters are not rational functions of Mandelstam variables.  First, we have 5 letters that are square roots of K\"all\'en functions\footnote{The letters $W_{i \in [118,122]} = r_{i-117}$ are special cases because we could have chosen $r_i^2$ as independent symbol letters instead of $r_i$, in which case we would have classified them as polynomial in Mandelstam variables rather than algebraic. However, because the $r_i^2$ are not (permutations of) cluster variables of $\Fl_{2,4;6}$ but we do obtain the $r_i$ from infinite mutation sequences, we consider it more suitable to think of these letters as algebraic.}:
\begin{align}
\begin{aligned}
    W_{118} &= \sqrt{v_1^2{+}v_3^2{+}v_5^2{-}2 v_1 v_3{-}2 v_1 v_5{-}2 v_3 v_5} \equiv r_1
    \,,
    &
    W_{119} &= T W_{118} \equiv r_2
    \,,
    \\
    W_{120} &= \sqrt{(v_7+v_9)^2-4 v_1 v_4} \equiv r_3\,,
    &
    W_{i\in[121,122]} &= T^{i-120} W_{120}
    \,,
\end{aligned}
\end{align}
and we will denote $r_4 = T r_3$, $r_5 = T r_4$. Next, there are 15 letters of mass-dimension four that cannot be written as polynomials in Mandelstam variables but are polynomial in spinor helicity variables:
\begin{align}
\begin{aligned}
    W_{123}
    &= \epsilon_{1234}
    \,,
    &W_{i\in [124,128]} &= T^{i{-}123} W_{123}
    \,,
    \\
    W_{129}
    &= \epsilon_{1235}
    \,,
    &W_{i\in [130,134]} &= T^{i{-}129} W_{129}
    \,,
    \\
    W_{135}
    &= \epsilon_{1245}
    \,,
    &W_{i\in [136,137]} &= T^{i{-}135} W_{135}
    \,,
\end{aligned}\label{eq:type5}
\end{align}
and there are 19 similar letters of mass-dimension six:
\begin{align}
\begin{aligned}
    W_{138}
    &= \Delta_6
    \,,
    \\
    W_{139}
    &= v_7 \epsilon_{5612}{-}v_1 \epsilon_{4561}
    \,,
    & W_{i\in [140,144]} &= T^{i{-}139} W_{139}
    \,,
    \\
    W_{145}
    &= v_5 \epsilon_{1234}{-}v_3 \epsilon_{1245}{-}(v_5{-}v_8) \epsilon_{2345}{+}(v_3{-}v_8) \epsilon_{3451}
    \,,
    & W_{i\in [146,150]} &= T^{i{-}145} W_{145}
    \,,
    \\
    W_{151}
    &= v_8\epsilon_{6123}{-}v_6 \epsilon_{1234}
    \,,
    & W_{i\in [152,156]} &= T^{i{-}151} W_{151}
    \,.
\end{aligned}\label{eq:type6}
\end{align}

The next 25 letters are algebraic functions of Mandelstam variables having the form $(-r_i+ [\text{polynomial in }v])/(r_i + [\text{polynomial in }v])$:
\begin{align}
\begin{aligned}
    W_{157}
    &= \frac{-r_1+v_1+v_3-v_5}{r_1+v_1+v_3-v_5}
    \,,
    & W_{i\in [158,160]} &= T^{i{-}157} W_{157}
    \,,
    \\
    W_{161}
    &= \frac{-r_1+v_1-v_3+v_5-2 v_7}{r_1+v_1-v_3+v_5-2 v_7}
    \,,
    & W_{i\in [162,166]} &= T^{i{-}161} W_{161}
    \,,
    \\
    W_{167}
    &= \frac{-r_3+v_7+v_9}{r_3+v_7+v_9}
    \,,
    & W_{i\in [168,169]} &= T^{i{-}167} W_{167}
    \,, \label{eq:algebraicone}
    \\
    W_{170}
    &= \frac{-r_3+v_7-v_9}{r_3+v_7-v_9}
    \,,
    & W_{i\in [171,172]} &= T^{i{-}170} W_{170}
    \,,
    \\
    W_{173}
    &= \frac{-r_3-2 v_1+v_7+v_9}{r_3-2 v_1+v_7+v_9}
    \,,
    & W_{i\in [174,175]} &= T^{i{-}173} W_{173}
    \,,
    \\
    W_{176}
    &= \frac{-r_3+2 v_3-2 v_5+v_7-v_9}{r_3+2 v_3-2 v_5+v_7-v_9}
    \,,
    & W_{i\in [177,181]} &= T^{i{-}176} W_{176}\,.
\end{aligned}
\end{align}
Similarly, there are 54 letters of the form $(-\epsilon_{ijkl}+ [\text{polynomial in }v])/(\epsilon_{ijkl}
+ [\text{polynomial in }v])$ that are algebraic in terms of Mandelstam variables but ratios of polynomials in spinor helicity variables:
\begin{align}
    W_{182}
    &= \frac{{-}\epsilon_{1234}{+}v_1 v_2{-}v_3 v_2{+}v_5 v_2{+}v_3 v_7{-}v_1 v_8{-}v_7 v_8}{\epsilon_{1234}{+}v_1 v_2{-}v_3 v_2{+}v_5 v_2{+}v_3 v_7{-}v_1 v_8{-}v_7 v_8}
    \,,
    &W_{i\in [183,187]} = T^{i{-}182} W_{182}
    \,,
    \nonumber\\
    W_{188}
    &=
    W_{182}\vert_{v_5 \to -v_5}
    \,,
    &W_{i\in [189,190]} = T^{i{-}188} W_{188}
    \,,
    \nonumber\\
    W_{194}
    &=
    W_{188}\vert_{v_2 \to -v_2, v_8 \to -v_8}
    \,,
    &W_{i\in [195,199]} = T^{i{-}194} W_{194}
    \,,
    \nonumber\\
    W_{200}
    &=
    W_{194}\vert_{v_2 \to -v_2, v_5 \to -v_5}
    \,,
    &W_{i\in [201,205]} = T^{i{-}200} W_{200}
    \,,
    \nonumber\\
    W_{206}
    &= \frac{{-}\epsilon_{1234}{+}v_2 (v_1 {-}v_3 {-}v_5 {+}2 v_8) {+}v_3 v_7{+}v_1 v_8{-}v_7 v_8}{\epsilon_{1234}{+}v_2 (v_1 {-}v_3 {-}v_5 {+}2 v_8) {+}v_3 v_7{+}v_1 v_8{-}v_7 v_8}
    \,,
    &W_{i\in [207,211]} = T^{i{-}206} W_{206}
    \,,
    \nonumber\\
    W_{218}
    &=
    \frac{{-}\epsilon_{1234}{+}v_1(v_2{+}2v_5{-}2v_7{-}v_8){+}v_2(v_5{-}v_3){+}v_7 (v_3{-}v_8)}{\epsilon_{1234}{+}v_1(v_2{+}2v_5{-}2v_7{-}v_8){+}v_2(v_5{-}v_3){+}v_7 (v_3{-}v_8)}
    \,,
    &W_{i\in [219,220]} = T^{i{-}218} W_{218}
    \,,
    \nonumber\\
    W_{224}
    &= \frac{{-}\epsilon_{1234}{+} v_1(v_2 {+} 2(v_1{-}v_3{-}v_7) {+} v_8) {+}v_2 (v_5{-}v_3){+}v_7(v_3 {-} v_8)}{\epsilon_{1234}{+} v_1(v_2 {+} 2(v_1{-}v_3{-}v_7) {+} v_8) {+}v_2 (v_5{-}v_3){+}v_7(v_3 {-} v_8)}
    \,,
    &W_{i\in [225,229]} = T^{i{-}224} W_{224}
    \,,
    \nonumber\\
    W_{230}
    &= \frac{\epsilon_{1234}{+}v_2 (v_5{-}v_3) {+} v_1 (v_2{+}2 v_5{-}v_8) {+} v_7 (v_3{-}2 v_5{+}v_8)}{{-}\epsilon_{1234}{+}v_2 (v_5{-}v_3) {+} v_1 (v_2{+}2 v_5{-}v_8) {+} v_7 (v_3{-}2 v_5{+}v_8)}
    \,,
    &W_{i\in [231,235]} = T^{i{-}230} W_{230}
    \,,
    \nonumber\\
    W_{236}
    &= \frac{\epsilon_{1234}{+} v_2(v_3{+}v_5) {+} (2v_5{-}v_7)(v_3-v_8) {+} v_1(v_8{-}v_2)}{{-}\epsilon_{1234}{+} v_2(v_3{+}v_5) {+} (2v_5{-}v_7)(v_3-v_8) {+} v_1(v_8{-}v_2)}
    \,,
    & W_{i\in [237,241]} = T^{i{-}236} W_{236}
    \,,
    \nonumber\\
    W_{242}
    &= \frac{{-}\epsilon_{1235}{-}v_1 (v_4{+}v_6{-}v_8){+}v_2 (v_3{+}v_5{-}v_9){+}v_7 (v_6{-}v_8{+}v_9{-}v_3)}{\epsilon_{1235}{-}v_1 (v_4{+}v_6{-}v_8){+}v_2 (v_3{+}v_5{-}v_9){+}v_7 (v_6{-}v_8{+}v_9{-}v_3)}
    \,,
    & W_{i\in [243,247]} = T^{i{-}242} W_{242}
    \,.
    \label{eq:type4}
\end{align}
Finally, there are 10 letters of the form $(-r_i \epsilon_{jklm}+ [\text{polynomial in }v])/(r_i \epsilon_{jklm}
+ [\text{polynomial in }v])$ that, like those in~(\ref{eq:algebraicone}), are algebraic even in spinor helicity variables:
\begin{align}\begin{aligned}
    W_{275}
    & =
    \frac{{-}r_1 \epsilon_{1234} + w_{275}}{r_1 \epsilon_{1234} + w_{275}}
    \,,
    &W_{i\in [276,278]} &= T^{i{-}275} W_{275}
    \,,
    \\
    W_{281}
    & = \frac{{-}r_4 \epsilon_{1234} + w_{281}}{r_4 \epsilon _{1234} + w_{281}}
    \,,
    &W_{i\in [282,286]} &= T^{i{-}281} W_{281}
    \,,
\end{aligned}\end{align}
where
\begin{align}
\begin{aligned}
w_{275} &= (v_2{-}v_8) v_1^2{+}((v_5{+}v_7) v_8{+}v_3 (v_7{+}v_8{-}2 v_5){-}2 v_2 (v_3{+}v_5)) v_1{+}v_2 (v_3{-}v_5)^2{-}(v_3{-}v_5) v_7 (v_3{-}v_8)\,,\\
w_{281} &= (v_2{-}v_8) v_1^2{+}((v_5{+}v_7) v_8{+}v_3 (v_7{+}v_8{-}2 v_5){-}2 v_2 (v_3{+}v_5)) v_1{+}(v_3{-}v_5) (v_2 (v_3{-}v_5){+}v_7 (v_8{-}v_3))\,.
\end{aligned}
\end{align}


\subsection{\texorpdfstring{The $\mathcal{F\ell}_{2,4;6}$ cluster algebra}{The Fl2,4;6 cluster algebra}}

\begin{figure}[t]
     \centering
\adjustbox{scale=0.7}{\begin{tikzcd}
 	\boxed{P_{5,6}} & {\boxed{P_{1,6}}} & {\boxed{P_{1,2}}} \\
 	\boxed{\makecell[r]{P_{4,5}P_{1,2,3,6}\\-P_{4,6}P_{1,2,3,5}\\+P_{5,6}P_{1,2,3,4}}} & {\makecell[r]{P_{1,5}P_{1,2,3,6}\\-P_{1,6}P_{1,2,3,5}}} && {P_{1,2,3,6}} \\
        \boxed{P_{3,4,5,6}} & {P_{1,4,5,6}} & {P_{1,2,5,6}} \\
 	\boxed{P_{2,3,4,5}} & {P_{1,3,4,5}} & {P_{1,2,4,5}} & {P_{1,2,3,5}}\\
    &&& {\boxed{P_{1,2,3,4}}}\\
        \arrow[from=2-1, to=2-2]
        \arrow[from=2-2, to=2-4]
        \arrow[from=3-1, to=3-2]
        \arrow[from=3-2, to=3-3]
        \arrow[from=4-1, to=4-2]
        \arrow[from=4-2, to=4-3]
        \arrow[from=4-3, to=4-4]
        \arrow[from=4-4, to=5-4]
        \arrow[from=1-2, to=2-2]
        \arrow[from=2-2, to=3-2]
        \arrow[from=3-2, to=4-2]
        \arrow[from=1-3, to=3-3]
        \arrow[from=3-3, to=4-3]
        \arrow[from=2-4, to=4-4]
        \arrow[from=2-2, to=1-1]
        \arrow[from=2-4, to=1-2]
        \arrow[from=3-2, to=2-1]
        \arrow[from=3-3, to=2-2]
        \arrow[from=4-2, to=3-1]
        \arrow[from=4-3, to=3-2]
        \arrow[from=4-4, to=3-3]
        \arrow[from=2-2, to=1-3]
\end{tikzcd}}
\qquad\qquad
\adjustbox{scale=0.7}{\begin{tikzcd}
 	\boxed{\la 56 \ra} & {\boxed{\la 16 \ra}} & {\boxed{\la 12 \ra}} \\
 	\boxed{\makecell[r]{\la 45 \ra [45]\\+\la 46\ra [46]\\+\la 56\ra [56]}} & {\makecell[r]{\la 15 \ra [45]\\+\la 16 \ra [46]}} && {[45]} \\
        \boxed{[12]} & {[23]} & {[34]} \\
 	\boxed{[16]} & {-[26]} & {[36]} & {-[46]}\\
    &&& {\boxed{[56]}}\\
        \arrow[from=2-1, to=2-2]
        \arrow[from=2-2, to=2-4]
        \arrow[from=3-1, to=3-2]
        \arrow[from=3-2, to=3-3]
        \arrow[from=4-1, to=4-2]
        \arrow[from=4-2, to=4-3]
        \arrow[from=4-3, to=4-4]
        \arrow[from=4-4, to=5-4]
        \arrow[from=1-2, to=2-2]
        \arrow[from=2-2, to=3-2]
        \arrow[from=3-2, to=4-2]
        \arrow[from=1-3, to=3-3]
        \arrow[from=3-3, to=4-3]
        \arrow[from=2-4, to=4-4]
        \arrow[from=2-2, to=1-1]
        \arrow[from=2-4, to=1-2]
        \arrow[from=3-2, to=2-1]
        \arrow[from=3-3, to=2-2]
        \arrow[from=4-2, to=3-1]
        \arrow[from=4-3, to=3-2]
        \arrow[from=4-4, to=3-3]
        \arrow[from=2-2, to=1-3]
\end{tikzcd}}
\caption{The initial cluster of $\mathcal{F\ell}_{2,4;6}$ with nodes labeled by the Pl\"ucker coordinates of the partial flag variety (left) or by the corresponding spinor helicity variables (right), with frozen nodes boxed.}
    \label{figure: initial cluster 246}
\end{figure}
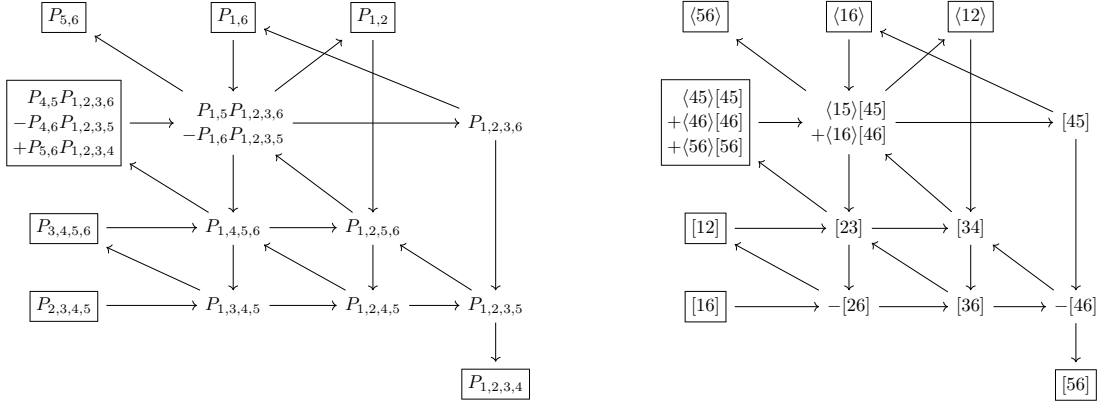

In this section, we review the cluster algebra associated to the partial flag variety $\mathcal{F}\ell_{2,4;6}$ that describes massless six-point kinematics.
Analogous to the five-point construction of Section~\ref{sec:fl235},  a point in $\mathcal{F\ell}_{2,4;6}$ is represented by a $4\times 6$ matrix $M$.
The Pl\"ucker coordinates of $\mathcal{F\ell}_{2,4;6}$ are $2\times 2$ and $4\times 4$ minors of $M$.
The relation between the Pl\"ucker coordinates of $\mathcal{F\ell}_{2,4;6}$ and the 6-point spinor helicity brackets is similar to the five-point case:
\begin{equation}
    \langle ij \rangle \leftrightarrow P_{i,j}\quad \text{and}\quad [ij]\leftrightarrow (-1)^{i+j+1}
    P_{k,l,m,n}\,,
    \quad
    \{k,l,m,n\} = [1,6] \setminus \{i,j\}
    \label{eq: 6pt map}
\end{equation}
for $i<j$ and $k<l<m<n$. The initial cluster of the $\mathcal{F\ell}_{2,4;6}$ cluster algebra is shown in Figure~\ref{figure: initial cluster 246}.
Unlike $\mathcal{F\ell}_{2,3;5}$, this cluster algebra is not of finite type; it has infinitely many cluster variables.
This is not a surprise due to its close relationship with the infinite $\text{Gr}(4,8)$ cluster algebra.  The embedding of the partial flag variety Pl\"ucker coordinates into those of $\Gr(4,8)$, which we denote by $\langle ijkl \rangle$, is
\begin{align}
    P_{i,j} \mapsto \langle ij78 \rangle\,,
    \quad\text{and}\quad
    P_{i,j,k,l}\mapsto \langle ijkl\rangle\,.
    \label{eq:embedding}
\end{align}
Under this embedding, the initial cluster of $\mathcal{F\ell}_{2,4;6}$ can be obtained from the initial cluster of $\Gr(4,8)$ by a sequence of 3 mutations, freezing 1 node, and deleting 3 nodes, as explained in~\cite{Bossinger:2024apm}.
The initial cluster of $\mathcal{F\ell}_{2,4;6}$ is given in Figure~\ref{figure: initial cluster 246}.  Note that every cluster variable of $\Fl_{2,4;6}$ maps into a cluster variable of $\Gr(4,8)$ under~(\ref{eq:embedding}), but the converse is of course not true.

Although there are infinitely many cluster variables, for the purpose of describing the six-point symbol alphabet we only need a finite subset.  We now tabulate the 74 cluster variables that we find relevant for this purpose in the next two sections.
First, we have all 15 two-entry Pl\"ucker coordinates $P_{i,j}$ and all 15 four-entry Pl\"ucker coordinates $P_{i,j,k,l}$, which correspond respectively to the spinor helicity brackets $\langle i j \rangle$ and $[ij]$.
There are 4 cluster variables quadratic in Pl\"uckers that take the form $Q_{i,j|k}= (-1)^{i+k+1} \sgn_{i,k} P_{i,k} P_{[1,6]\setminus \{i,k\}} + \{i \leftrightarrow j\} \leftrightarrow \la ik \ra [ik]+ \la jk \ra [jk]$
where $\sgn_{i,k}:=\sgn(k-i)$; these are
\begin{equation}
\begin{aligned}
    &Q_{1,2|6}  \leftrightarrow \la 16\ra [16]{+}\la26\ra [26]\,,\quad  Q_{4,5|6} \leftrightarrow \la 46\ra [46]{+} \la56\ra [56]\,,\\
    & Q_{2,3|1} \leftrightarrow \la 12\ra [12]{+}\la13\ra [13]\,,\quad Q_{5,6|1} \leftrightarrow \la 16\ra [16]{+} \la15\ra [15]\,,
\end{aligned}\label{eq:Qdef}
\end{equation}
and another 30 of the form
\begin{equation}
\begin{aligned}
    & R_{i,j|k_1,k_2} =(-1)^{j+k_1+1}\sgn_{j,k_1}P_{i,k_1}P_{[1,6] \setminus \{j,k_1\}} + \{k_1 \leftrightarrow k_2\} \leftrightarrow \la i k_1 \ra [jk_1] + \la i k_2 \ra [jk_2]
    \,,
\end{aligned}
\end{equation}
where $i,j\in [1,6],\,i\neq j$ and $k_1, k_2$ are the two smallest integers in the set $[1,6]\setminus \{i,j\}$.
Then there is a quadratic polynomial of the form
$C_{i,j,k}=(-1)^{i+j+1} \sgn_{i,j} P_{i,j} P_{[1,6]\setminus \{i,j\}} + \text{cyclic} \leftrightarrow \la ij\ra [ij] +\text{cyclic}$:
\begin{equation}
    C_{1,2,3}\leftrightarrow \la 12 \ra [12] + \la 13 \ra [13] + \la 23\ra [23]\,.
\end{equation}
Finally, we have 9 cubic polynomials of the form
\begin{align}
     U_{i,j|k,l} &= (-1)^{k+l+1}\text{sgn}_{k,l} P_{i,k}P_{j,l}P_{[1,6]\setminus \{k,l\} }+\{k \rightarrow j\} - (-1)^{m+n+1}\text{sgn}_{m,n} P_{i,j}P_{m,n}P_{[1,6]\setminus \{m,n\} }
     \notag \\
     & \leftrightarrow \la ik \ra\la jl \ra[kl]+ \la ij \ra\la jl \ra [jl]-\la ij\ra \la mn \ra[mn]\,,
\end{align}
where $\{m,n\}=[1,6]\setminus\{i,j,k,l\}$. Specifically, the 9 cubic polynomials are
\begin{align}
\begin{aligned}
     U_{1,3|2,4}\,, U_{1,4|2,3}\,, U_{2,3|1,4}\,, U_{6,3|5,4}\,, U_{6,4|5,3}\,, \overline{U_{1,2|3,4}}\,, \overline{U_{2,3|1,4}}\,, \overline{U_{2,4|1,3}}\,, \overline{U_{5,3|6,4}}\,,
\end{aligned}
\end{align}
where $\overline{X} =X\vert_{[\bullet] \leftrightarrow \la \bullet \ra}$
denotes parity conjugation.

Just like in the five-point case, we consider the closure of this set of cluster variables under the $S_6$ permutation group on the particle labels. In total this produces a set of 910 variables as follows: 15 distinct $P_{i,j}$, 15 distinct $P_{i,j,k,l}$, 60 distinct $Q_{i,j|k}$, 90 distinct $R_{i,j|k_1,k_2}$, 10 distinct $C_{i,j,k}$, 360 distinct $U_{i,j|k,l}$ and 360 distinct $\overline{U_{i,j|k,l}}$.
It may seem surprising that we need to consider the full $S_6$ permutation group in order to describe an alphabet for planar six-point integrals, but the reason will become more apparent shortly, and in any case we have already seen in the previous section that this was necessary for $n=5$.

\subsection{\texorpdfstring{Rational letters from $\Fl_{2,4;6}$ cluster variables}{Rational letters from Fl2,4;6 cluster variables}}\label{sec: 6-pt rational alphabet construction}

In this section we explore the connection between the 205 six-point letters that are rational functions of spinor helicity variables and the 910 $\Fl_{2,4;6}$ cluster variables defined in the previous section. Throughout this section, when we say ``cluster variable'' we mean up to $S_6$ permutations, and when we discuss symbol letters, we always mean up to overall sign.

We first look at the 117 letters that are polynomial in Mandelstam variables.  After expressing them in terms of spinor helicity variables, we find that 87 of them are cluster variables or products of cluster variables.
For the letters (\ref{eq:type1}) that are linear in Mandelstam variables we find
\begin{align}\begin{aligned}
    W_1 &= \la 12 \ra[12]
    \leftrightarrow P_{1,2}P_{3,4,5,6}
    \,,
    &W_7 &= \la 12 \ra[12]{+}\la 13 \ra[13]{+}\la 23 \ra[23]
    \leftrightarrow C_{1,2,3}\,,
    \\
    W_{10} &=-\la12\ra [12]{-}\la32\ra [32] \leftrightarrow -Q_{1,3|2}\,,
    &W_{16} &= {-}\la13\ra [13] {-} \la 23 \ra [23]
    \leftrightarrow -Q_{1,2|3}\,,\\
    W_{22} &= {-} \la16\ra [16] {-} \la 26 \ra [26]
    \leftrightarrow -Q_{1,2|6}\,,
    &W_{28} &= \la 13 \ra [13]
    \leftrightarrow {-}P_{1,3}P_{2,4,5,6}\,,\\
    W_{34} &= \la 53 \ra [53] {+} \la 63 \ra [63]
    \leftrightarrow Q_{5,6|3}\,,
    &W_{40} &= \la 36 \ra [36] {+} \la 46 \ra [46]
    \leftrightarrow Q_{3,4|6}\,,\\
    W_{46} &= \la 36 \ra [36]
    \leftrightarrow P_{3,6}P_{1,2,4,5}\,.
\end{aligned}\end{align}
Note that in~(\ref{eq:Qdef}) the arguments of $Q_{i,j|k}$ are always cyclically adjacent, but here we require $Q$'s that are out of order; this is the quickest way to see why we consider the closure of the set of cluster variables under all permutations. For the letters (\ref{eq:type2}) that are quadratic in Mandelstam variables we find
\begin{align}
\begin{aligned}
    W_{49} &= (\la61\ra [31]{+}\la62\ra [32])(\la31\ra [61]{+}\la32\ra [62]) \leftrightarrow R_{6,3|1,2} \overline{R_{6,3|1,2}}\,,\\
    W_{76}&=(\la41\ra[61]{+}\la42\ra[62])(\la61\ra[41]{+}\la62\ra[42]) \leftrightarrow R_{4,6|1,2}\overline{R_{4,6|1,2}}\,,
    \\
    W_{82}&=(\la61\ra[21]{+}\la63\ra[23])(\la21\ra[61]{+}\la23\ra[63]) \leftrightarrow R_{6,2|1,3}\overline{R_{6,2|1,3}}\,,
    \\
    W_{88} &= (\la51\ra [61]{+}\la52\ra [62])(\la61\ra [51]{+}\la62\ra [52]) \leftrightarrow R_{5,6|1,2} \overline{R_{5,6|1,2}}\,,\\
    W_{94} &= (\la21\ra [31]{+}\la24\ra [34])(\la31\ra [21]{+}\la34\ra [24]) \leftrightarrow R_{2,3|1,4}\overline{R_{2,3|1,4}}\,.
\end{aligned}
\end{align}
Finally for the letters (\ref{eq:type3}) that are cubic in Mandelstam variables we find
\begin{align}
\begin{aligned}
    W_{106}&=(\la12\ra\la34\ra[24]{+}\la13\ra\la34\ra[34]{-}\la13\ra\la56\ra[56])(\la24\ra[12][34]{+}\la34\ra[13][34]{-}\la56\ra[13][56])\\
    &\leftrightarrow U_{1,3|2,4}\overline{U_{1,3|2,4}}\,,\\
    W_{112}&= W_{106} \rvert_{i \to 2 - i\,\text{mod}\,6} \leftrightarrow U_{1,5|6,4}\overline{U_{1,5|6,4}}\,.
\end{aligned}
\end{align}
The remaining 30 letters $W_{[52,75] \cup [100,105]}$ that are polynomial in Mandelstam variables are not cluster variables of $\Fl_{2,4;6}$ (or products thereof).

Next we consider the 34 letters $W_{[123,156]}$ shown in (\ref{eq:type5}) and (\ref{eq:type6}) that are not polynomial in Mandelstam variables but are polynomial when expressed in terms of spinor helicity variables.  We find that none of these are cluster variables of $\Fl_{2,4;6}$ (or products thereof).

Finally we consider the 54 letters (\ref{eq:type4}) that are rational functions only when expressed in terms of spinor helicity variables.  We find that
\begin{align}
    W_{182}&=\frac{(\la42\ra [12]+\la43\ra [13])\la12\ra [24]}{(\la12\ra [42]+\la13\ra [43])\la24\ra [12]}\leftrightarrow -\frac{R_{4,1|2,3}P_{1,2}P_{1,3,5,6}}{\overline{R_{4,1|2,3}}P_{2,4}P_{3,4,5,6}}\,,
    \nonumber\\
    W_{188}&=\frac{(\la42\ra [12]+\la43\ra [13])\la13\ra [34]}{(\la12\ra [42]+\la13\ra [43])\la34\ra [13]}\leftrightarrow -\frac{R_{4,1|2,3}P_{1,3}P_{1,2,5,6}}{\overline{R_{4,1|2,3}}P_{3,4}P_{2,4,5,6}}\,,
    \nonumber\\
    W_{194}&=\frac{(\la21\ra [41]+\la23\ra [43])\la34\ra [23]}{(\la41\ra [21]+\la43\ra [23])\la23\ra [34]}\leftrightarrow \phantom{-} \frac{R_{2,4|1,3}P_{3,4}P_{1,4,5,6}}{\overline{R_{2,4|1,3}}P_{2,3}P_{1,2,5,6}}\,,
    \nonumber\\
    W_{200}&=\frac{(\la32\ra [12]+\la34\ra [14])\la12\ra [23]}{(\la12\ra [32]+\la14\ra [34])\la23\ra [12]}\leftrightarrow \phantom{-} \frac{R_{3,1|2,4}P_{1,2}P_{1,4,5,6}}{\overline{R_{3,1|2,4}}P_{2,3}P_{3,4,5,6}}\,,
    \nonumber\\
    W_{206}&=\frac{(\la23\ra [13]+\la24\ra [14])\la13\ra [23]}{(\la13\ra [23]+\la14\ra [24])\la23\ra [13]}\leftrightarrow -\frac{R_{2,1|3,4}P_{1,3}P_{1,4,5,6}}{\overline{R_{2,1|3,4}}P_{2,3}P_{2,4,5,6}}\,,
    \\
    W_{218}&=\frac{(\la41\ra [21]+\la43\ra [23])\la12\ra [14]}{(\la21\ra [41]+\la23\ra [43])\la14\ra [12]}\leftrightarrow \phantom{-} \frac{R_{4,2|1,3}P_{1,2}P_{2,3,5,6}}{\overline{R_{4,2|1,3}}P_{1,4}P_{3,4,5,6}}\,,
    \nonumber\\
    W_{224}&=\frac{(\la31\ra [21]+\la34\ra [24])\la12\ra [13]}{(\la21\ra [31]+\la24\ra [34])\la13\ra [12]}\leftrightarrow - \frac{R_{3,2|1,4}P_{1,2}P_{2,4,5,6}}{\overline{R_{3,2|1,4}}P_{1,3}P_{3,4,5,6}}\,,\nonumber\\
    W_{230}&= \frac{ (\la 12 \ra \la 34 \ra [24] {+} \la 13 \ra \la 34 \ra [34] {-} \la 13 \ra \la 56 \ra [56]) [13] }{( [12] [34] \la 24 \ra {+}  [13] [34] \la 34 \ra {-} [13] [56] \la 56 \ra) \la 13 \ra  }\leftrightarrow-\frac{ U_{1,3|2,4} P_{2,4,5,6}  } { \overline{U_{1,3|2,4}} P_{1,3}  }\,,\nonumber\\
    W_{236}&= W_{230} \rvert_{i \to 5-i\,\text{mod}\,6}\leftrightarrow -\frac{ U_{4,2|3,1} P_{1,3,5,6} } { \overline{U_{4,2|3,1}} P_{2,4}  }\,,\nonumber
\end{align}
but the remaining 6 letters of this type, $W_{[242,247]}$, are not ratios of products of $\Fl_{2,4;6}$ cluster variables.

Two important comments are in order regarding the 70 letters $W_{[52,75] \cup [100,105]\cup [123,156] \cup [242,247]}$ that are polynomials or rational functions of spinor helicity variables but are not (ratios of products of) cluster variables. Firstly, although we have generated a large number of $\Fl_{2,4;6}$ cluster variables by performing various sequences of mutations on the initial cluster, one might wonder whether some of the missing letters involve cluster variables that just happen to lie beyond the horizon of our calculation. This possibility can be excluded by comparing (all permutations of) each letter, when expressed in terms of $\Gr(4,8)$ Pl\"ucker coordinates using~(\ref{eq:embedding}), against the compete set of $\Gr(4,8)$ cluster variables of the same degree\footnote{The maximum degree polynomial that arises when the planar two-loop six-point alphabet is embedded into $\Gr(4,8)$ is six. The $\Gr(4,8)$ cluster algebra has respectively 70, 120, 174, 208, 296, 304 cluster variables of degree 1 through 6 in Pl\"ucker coordinates; see for example~\cite{Cheung:2022itk}. We thank J.-R.~Li for providing data on these cluster variables.}. We found this check rather trivial for the most complicated (sextic) letters $W_{[100,105] \cup [138,156]}$ since each has homogeneous degree three in all 8 of the Pl\"ucker indices, but $\Gr(4,8)$ has no such cluster variables.

Secondly, we have further checked that no product of powers of the 70 missing letters can be expressed as ratios of products of cluster variables, so the absence of fully 70 letters is not merely an artifact of having chosen a non-optimal basis for the symbol alphabet.


\subsection{Algebraic letters from infinite mutation sequences}

In the context of SYM theory, where the cluster algebra relevant for $n$-point scattering is $\Gr(4,n)$, two features emerge starting at $n \ge 8$: the cluster algebra is infinite, and amplitudes have symbol letters that are algebraic, not polynomial, in Pl\"ucker coordinates.  Several approaches for naturally associating algebraic symbol letters to infinite cluster algebras have been explored in the physics literature~\cite{Drummond:2019cxm,Arkani-Hamed:2019rds,Henke:2019hve,Herderschee:2020lgb,Ren:2021ztg}, and a construction of this type is known in the mathematics literature; see for example~\cite{ccanakcci2018cluster}. In this section we show that for the infinite cluster algebra $\Fl_{2,4;6}$, the construction of~\cite{Drummond:2019cxm} based on infinite mutation sequences produces (after considering suitable permutations of cluster variables) \emph{all} 40 algebraic letters of the planar two-loop six-point alphabet; one example of this construction was demonstrated already in~\cite{2638233}.

\begin{figure}[t]
    \centering
    \adjustbox{scale=.8}{\begin{tikzcd}
        & & \boxed{f_z}\\
        & & {z_0}\\
        {b_1} & {b_2} && {b_3}\\
        & & {w_0}\\
        & \boxed{f_{w_1}} & \boxed{f_{w_2}} & \boxed{f_{w_3}}\\
        \arrow[from=1-3, to=2-3]
        \arrow[from=2-3, to=3-1]
        \arrow[from=2-3, to=3-2]
        \arrow[from=2-3, to=3-4]
        \arrow[from=3-1, to=4-3]
        \arrow[from=3-2, to=4-3]
        \arrow[from=3-4, to=4-3]
        \arrow[from=4-3, to=5-2]
        \arrow[from=4-3, to=5-3]
        \arrow[from=4-3, to=5-4]
        \arrow[from=4-3, to=2-3,shift left]
        \arrow[from=4-3, to=2-3,shift right]
    \end{tikzcd}}
\caption{The starting point of the infinite mutation sequence construction described in~\cite{Drummond:2019cxm}. The boxed nodes are frozen and nodes not connected to $z_0$ or $w_0$ are omitted.}
\label{fig: Gr48 infinite start}
\end{figure}
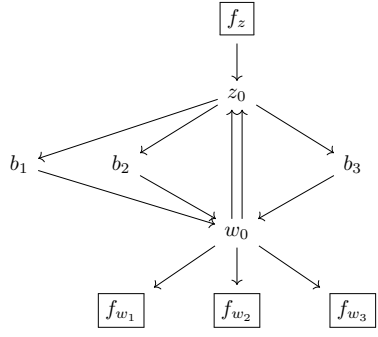

\begin{figure}[t]
    \centering
    \adjustbox{scale=.8}{\begin{tikzcd}
        & \\
        & {b} & \\
        {z_0} & & {w_0}\\
        \boxed{f_z} & & \boxed{f_w}\\
        \arrow[from=3-1, to=2-2]
        \arrow[from=2-2, to=3-3]
        \arrow[from=3-3, to=3-1, shift left]
        \arrow[from=3-3, to=3-1, shift right]
        \arrow[from=4-1, to=3-1]
        \arrow[from=3-3, to=4-3]
    \end{tikzcd}}
    $\longrightarrow$
    \adjustbox{scale=.8}{\begin{tikzcd}
        & \\
        & {b} & \\
        {z_0} & & {z_1}\\
        \boxed{f_z} & & \boxed{f_w}\\
        \arrow[from=2-2, to=3-1]
        \arrow[from=3-3, to=2-2]
        \arrow[from=3-1, to=3-3, shift left]
        \arrow[from=3-1, to=3-3, shift right]
        \arrow[from=4-1, to=3-1]
        \arrow[from=4-3, to=3-3]
    \end{tikzcd}}
    $\longrightarrow$
    \adjustbox{scale=.8}{\begin{tikzcd}
        & \\
        & {b} & \\
        {z_2} & & {z_1}\\
        \boxed{f_z} & & \boxed{f_w}\\
        \arrow[from=3-1, to=2-2]
        \arrow[from=2-2, to=3-3]
        \arrow[from=3-3, to=3-1, shift left]
        \arrow[from=3-3, to=3-1, shift right]
        \arrow[from=3-1, to=4-1]
        \arrow[from=4-3, to=3-3]
        \arrow[from=4-1, to=3-3,shift left]
        \arrow[from=4-1, to=3-3,shift right]
    \end{tikzcd}}
    $\longrightarrow~\cdots$
\caption{The infinite mutation sequence that starts from mutating on $w_0$. For clarity we omit arrows and nodes that are not connected to the two nodes in the first cluster where $z_0$ and $w_0$ sit.}
\label{fig: Gr48 infinite sequence}
\end{figure}
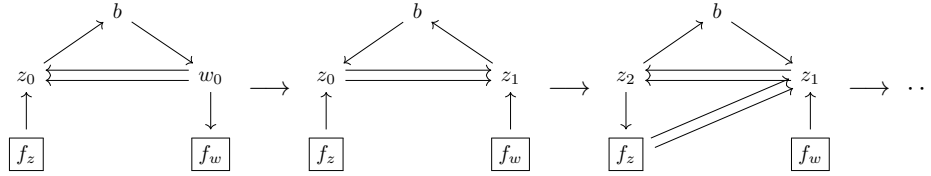

The infinite mutation sequence construction of~\cite{Drummond:2019cxm} starts with a cluster of the type shown in Figure~\ref{fig: Gr48 infinite start} and proceeds by alternately mutating on the nodes labeled $z_0$ and $w_0$, as depicted in Figure~\ref{fig: Gr48 infinite sequence}; then the sequence of cluster variables $z_1,z_2,\ldots$ that appears is given explicitly by
\begin{align}
    z_n=\frac{1}{2^{n+1}} \left[(z_0+B_z\sqrt{\Delta})(P+\sqrt{\Delta})^n+(z_0-B_z\sqrt{\Delta})(P-\sqrt{\Delta})^n\right],
    \label{eq: zn close form}
\end{align}
where
\begin{align}
    z_1=\frac{b_1 b_2 b_3 +f_w z_0^2}{w_0}
    \,,~
    P=\frac{f_z w_0+z_1}{z_0}
    \,,~
    \Delta=P^2-4 f_z f_w\,,
    ~
    f_w = f_{w_1} f_{w_2} f_{w_3}\,,
    ~
    B_z=\frac{2 z_1-z_0 P}{\Delta}\,.
    \label{eq: definition for zn}
\end{align}
The other infinite sequence, that starts by mutating on $z_0$, generates another set of cluster variables $w_n$ that is given by swapping $z$ and $w$ in~\eqref{eq: zn close form} and~\eqref{eq: definition for zn}.
The authors of~\cite{Drummond:2019cxm} noted that (in the case of $\Gr(4,8)$) the quantity $(z_0 + B_z \sqrt{\Delta})(z_0 - B_z \sqrt{\Delta})$ (and its $z \leftrightarrow w$ partner) is always a product of powers of cluster variables and suggested to consider the large $n$ limit of $z_n^2/(z_{2n}(z_0 + B_z \sqrt{\Delta})(z_0 - B_z \sqrt{\Delta}))$ as a new, multiplicatively independent letter.  Since the first term in~(\ref{eq: zn close form}) dominates in this limit, this amounts to associating the algebraic ratios
\begin{equation}
    \frac{z_0+B_z\sqrt{\Delta}}{z_0-B_z\sqrt{\Delta}} \quad \text{and its partner} \quad
    \frac{w_0+B_w\sqrt{\Delta}}{w_0-B_w\sqrt{\Delta}}
\end{equation}
to such an infinite mutation sequence. In the case of SYM theory, this construction precisely produces all
algebraic letters that are currently known to appear~\cite{He:2019jee} in the symbols of 8-point amplitudes.

\begin{figure}[t]
     \centering
\adjustbox{scale=0.6}{\begin{tikzcd}
    {\color{red}1\color{black}\ \boxed{P_{5,6}}}
    & {\color{red}2\color{black}\ \boxed{P_{1,6}}}
    & {\color{red}3\color{black}\ \boxed{P_{1,2}}} \\
    {\color{red}4\color{black}\ \boxed{\makecell[r]{P_{4,5}P_{1,2,3,6}\\-P_{4,6}P_{1,2,3,5}\\+P_{5,6}P_{1,2,3,4}}}}
    & {\color{red}8\color{black}\ \makecell[r]{P_{1,5}P_{1,2,3,6}\\-P_{1,6}P_{1,2,3,5}}}
    && {\color{red}9\color{black}\ P_{1,2,3,6}} \\
    {\color{red}5\color{black}\ \boxed{P_{3,4,5,6}}}
    & {\color{red}10\color{black}\ P_{1,4,5,6}}
    & {\color{red}11\color{black}\ P_{1,2,5,6}} \\
    {\color{red}6\color{black}\ \boxed{P_{2,3,4,5}}}
    & {\color{red}12\color{black}\  P_{1,3,4,5}}
    & {\color{red}13\color{black}\  P_{1,2,4,5}}
    & {\color{red}14\color{black}\  P_{1,2,3,5}}\\
    &&& {\color{red}7\color{black}\ \boxed{P_{1,2,3,4}}}\\
        \arrow[from=2-1, to=2-2]
        \arrow[from=2-2, to=2-4]
        \arrow[from=3-1, to=3-2]
        \arrow[from=3-2, to=3-3]
        \arrow[from=4-1, to=4-2]
        \arrow[from=4-2, to=4-3]
        \arrow[from=4-3, to=4-4]
        \arrow[from=4-4, to=5-4]
        \arrow[from=1-2, to=2-2]
        \arrow[from=2-2, to=3-2]
        \arrow[from=3-2, to=4-2]
        \arrow[from=1-3, to=3-3]
        \arrow[from=3-3, to=4-3]
        \arrow[from=2-4, to=4-4]
        \arrow[from=2-2, to=1-1]
        \arrow[from=2-4, to=1-2]
        \arrow[from=3-2, to=2-1]
        \arrow[from=3-3, to=2-2]
        \arrow[from=4-2, to=3-1]
        \arrow[from=4-3, to=3-2]
        \arrow[from=4-4, to=3-3]
        \arrow[from=2-2, to=1-3]
\end{tikzcd}}
$\quad  \quad$
\adjustbox{scale=0.6}{\begin{tikzcd}
    {\color{red}1\color{black}\ \boxed{f_{w_2}}}
    & {\color{lightgray}2\ \boxed{P_{1,6}}\color{black}}
    & {\color{red}3\color{black}\ \boxed{f_{w_3}}} \\
    {\color{lightgray}4\ \boxed{\makecell[r]{P_{4,5}P_{1,2,3,6}\\-P_{4,6}P_{1,2,3,5}\\+P_{5,6}P_{1,2,3,4}}}\color{black}}
    & {\color{lightgray}8\ \makecell[r]{P_{2,6}P_{1,2,3,5}\\-P_{1,6}P_{2,3,4,5}}\color{black}}
    && {\color{red}9\color{black}\ b_3} \\
    {\color{red}5\color{black}\ \boxed{f_{w_1}}}
    & {\color{red}10\color{black}\ w_0}
    & {\color{red}11\color{black}\ b_1} \\
    {\color{lightgray}6\ \boxed{P_{2,3,4,5}}\color{black}}
    & {\color{lightgray}12\  P_{1,3,4,5}\color{black}}
    & {\color{red}13\color{black}\  b_2}
    & {\color{red}14\color{black}\  z_0}\\
    &&& {\color{red}7\color{black}\ \boxed{f_z}}\\
        \arrow[from=1-3, to=2-2, color=lightgray, dashed]
        \arrow[from=1-3, to=3-2]
        \arrow[from=1-2, to=4-2, bend right=50, color=lightgray, dashed]
        \arrow[from=1-1, to=2-2, color=lightgray, dashed]
        \arrow[from=1-1, to=3-2, bend left=10]
        \arrow[from=2-1, to=4-3, bend right=20, color=lightgray, dashed]
        \arrow[from=3-1, to=3-2]
        \arrow[from=4-1, to=4-2, color=lightgray, dashed]
        \arrow[from=2-2, to=1-2, color=lightgray, dashed]
        \arrow[from=2-2, to=2-4, color=lightgray, dashed]
        \arrow[from=4-4, to=5-4]
        \arrow[from=4-4, to=3-2, shift left]
        \arrow[from=4-4, to=3-2, shift right]
        \arrow[from=3-2, to=3-3]
        \arrow[from=3-2, to=2-4]
        \arrow[from=3-2, to=4-3]
        \arrow[from=3-3, to=1-1, bend right=30, color=lightgray, dashed]
        \arrow[from=3-3, to=4-4]
        \arrow[from=4-2, to=2-2, bend right=60, color=lightgray, dashed]
        \arrow[from=2-4, to=1-3, color=lightgray, dashed]
        \arrow[from=2-4, to=1-1, color=lightgray, dashed]
        \arrow[from=2-4, to=3-1, color=lightgray, dashed]
        \arrow[from=2-4, to=4-4]
        \arrow[from=4-3, to=1-3, bend right=20, color=lightgray, dashed]
        \arrow[from=4-3, to=1-1, color=lightgray, dashed]
        \arrow[from=4-3, to=3-1, color=lightgray, dashed]
        \arrow[from=4-3, to=4-4]
\end{tikzcd}}
\caption{
Left: The initial cluster of $\mathcal{F\ell}_{2,4;6}$, with nodes labeled in red.
Right: The cluster obtained after performing the mutation sequence $10\rightarrow 8\rightarrow 9\rightarrow 13 \rightarrow 10$, highlighting the embedded cluster of the type shown in Figure~\ref{fig: Gr48 infinite start} with the nodes and edges not connected to $w_0$ or $z_0$ dimmed.}
\label{fig: labeled Fl246}
\end{figure}
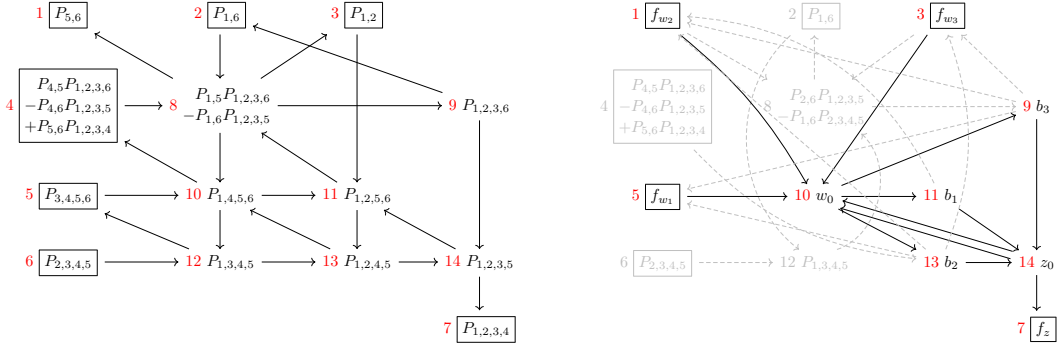

In order to apply this construction to $\Fl_{2,4;6}$, we note that beginning with the initial cluster shown in the left panel of Figure~\ref{fig: labeled Fl246}, the five-step mutation sequence
$10\rightarrow 8 \rightarrow 9\rightarrow 13\rightarrow 10$ (referring to the nodes as labeled in red in the figure) produces a cluster of the type shown in Figure~\ref{fig: Gr48 infinite start}, with the minor difference that all arrows are reversed.  The cluster variables sitting at the relevant nodes on the right panel of Figure~\ref{fig: labeled Fl246} are:
\begin{align}
\begin{aligned}
    z_0 &= P_{1,2,3,5} \leftrightarrow -[46]\,,
    & w_0 &= -R_{5,4|1,2} \leftrightarrow -\la51\ra[41]-\la52\ra[42]\,,\\
    f_z &= P_{1,2,3,4} \leftrightarrow [56]\,,
    & b_1 &= P_{1,2,5,6} \leftrightarrow [34]\,,\\
    f_{w_1} &= P_{3,4,5,6} \leftrightarrow [12]\,,
    & b_2 &= -R_{5,6|1,2} \leftrightarrow -\la51\ra[61]-\la52\ra[62]\,,\\
    f_{w_2} &= P_{5,6} \leftrightarrow \la56\ra\,,
    & b_3 &= -R_{3,4|1,2} \leftrightarrow  -\la31\ra[41] -\la32\ra[42]\,,\\
    f_{w_3} &= P_{1,2} \leftrightarrow \la12\ra\,.
\end{aligned}
\end{align}

In this case a short calculation reveals that $(z_0 + B_z \sqrt{\Delta})(z_0 - B_z \sqrt{\Delta})$ and its $z \leftrightarrow w$ partner are not products of (powers of) cluster variables, but have an extra denominator factor $r_1^2$.  Therefore it is natural to assign the algebraic letters
\begin{align}
\begin{aligned}
    K_z
    &=\frac{z_0+B_z\sqrt{\Delta}}{z_0-B_z\sqrt{\Delta}}\frac{1}{r_1^2}
    \quad \text{and} \quad
    K_w
    =
    \frac{w_0+B_w\sqrt{\Delta}}{w_0-B_w\sqrt{\Delta}}\frac{1}{r_1^2}
\end{aligned}
\end{align}
to the infinite mutation sequences that start from the cluster shown in the right panel of Figure~\ref{fig: labeled Fl246}.

As we did for the rational letters, we consider the union of $\{K_z, K_w\}$ under the $S_6$ permutation group and find that the resulting set contains all of the square roots $r_1,\ldots,r_5$ (among others).  A multiplicative basis for the subset involving the five $r_i$ is provided by
\begin{align}
    &L_{r_1, i\in[1,8]}=\left\{K_z,\, K_z|_{(13)(24)},\, K_z|_{(1324)},\, K_w,\, K_w|_{(3645)},\, K_w|_{(153)(264)},\, K_w|_{(154)(263)},\, K_w|_{(135246)}\right\},\notag \\
    &L_{r_2}=L_{r_1}|_{(123456)}
    ,\quad
    L_{r_3}=L_{r_1}|_{(465)}
    ,\quad
    L_{r_4}=L_{r_1}|_{(1234)}
    ,\quad
    L_{r_5}=L_{r_1}|_{(1356)(24)}\,, \label{eq:algebraicbasis}
\end{align}
where we use the cycle notation to represent elements of $S_6$ acting on the Pl\"ucker indices (which we emphasize is not the same as using~\eqref{eq: 6pt map} and then acting on the spinor helicity indices). We find that this subset provides a multiplicative basis for all algebraic letters of the massless two-loop six-point alphabet.  For example, the letters involving the root $r_1$ are given explicitly by
\begin{align}\label{eq:algLetterMap}
    (W_{118})^2&=r_1^2 = -\frac{L_{r_1,4} L_{r_1,6}}{L_{r_1,2} L_{r_1,3} L_{r_1,8}}
    \,,~
    &W_{157} &= \frac{L_{r_1,2}^2 L_{r_1,3} L_{r_1,8}}{L_{r_1,4}^2 L_{r_1,6}^2}
    \,,~
    &W_{159} &= \frac{L_{r_1,3} L_{r_1,8}}{L_{r_1,1} L_{r_1,4}}
    \,,\notag\\
    W_{161} &= \frac{L_{r_1,4}^2 L_{r_1,6} L_{r_1,7}}{L_{r_1,2} L_{r_1,3} L_{r_1,8}^2}
    \,,~
    &W_{163} &= \frac{L_{r_1,1} L_{r_1,4}^2 L_{r_1,6}^2}{L_{r_1,2}^2 L_{r_1,3} L_{r_1,5} L_{r_1,8}}
    \,,~
    &W_{165} &= \frac{L_{r_1,4}^2 L_{r_1,6}^2}{L_{r_1,2} L_{r_1,3} L_{r_1,8}^2}
    \,,\notag\\
    W_{275} &= \frac{L_{r_1,1}^2 L_{r_1,4}^6 L_{r_1,6}^5 L_{r_1,7}}{L_{r_1,2}^4 L_{r_1,3}^5 L_{r_1,8}^5}
    \,,~
    &W_{277} &= \frac{L_{r_1,2} L_{r_1,3} L_{r_1,8}^2}{L_{r_1,4} L_{r_1,5} L_{r_1,6} L_{r_1,7}}\,,
\end{align}
and the algebraic letters containing the other roots $r_2,\ldots,r_5$ be expressed in terms of the bases $\{L_{r_2,i}\},\ldots,\{L_{r_5,i}\}$ via similar formulas.

\section{Conclusion}

In this paper we have explored the connection between the complete 245-letter symbol alphabet of planar massless two-loop six-point Feynman integrals and cluster variables associated to the partial flag variety $\Fl_{2,4;6}$ via a construction that generalizes the five-point analysis of~\cite{Bossinger:2022eiy} and makes use of the embedding~\cite{Bossinger:2024apm} of the $\Fl_{2,4;6}$ cluster algebra into that of $\Gr(4,8)$.  Our findings are summarized in the table\footnote{More details are available in ancillary files attached to the arXiv submission and on GitHub: \github.}:
\begin{center}
\begin{tabular}{lrll}
Letters & \# & Description & Relation to $\Fl_{2,4;6}$\\ \hline
  $W_{[1,51]\cup[76,99]\cup[106,117]}$ & 87 & polynomial in MV & \cmark\, CV \\
   $W_{[52,75]\cup [100,105]}$ & 30 & polynomial in MV & \xmark\, not CV\\
   $W_{[118,122] \cup [157,181]\cup [275,278] \cup [281,286]}$ & 40 & algebraic in SHV & \cmark\, infinite  sequences\\
   $W_{[123,156]}$ & 34 & polynomial in SHV & \xmark\, not CV\\
     $W_{[182,190]\cup[194,211] \cup [218,220] \cup [224,241]}$ & 48 & ratio of polynomials in SHV & \cmark\, ratio of CV \\
     $W_{[242,247]}$ & 6 & ratio of polynomials in SHV & \xmark\, not ratio of CV
\end{tabular}

\end{center}
Here MV and SHV stand for ``Mandelstam variables'' and ``spinor helicity variables'', and CV is shorthand for the mouthful ``products of (permutations of) cluster variables, up to overall sign''.

Of the 70 symbol letters marked with an \ding{55} in the table above, 27 appear only at $\mathcal{O}(\epsilon)$ or higher in dimensional regularization and are therefore not relevant for suitably defined finite quantities in four dimensions; these letters are $W_{[129,137] \cup [139,156]}$.
An additional 7 letters are six-point analogues of the five-point letter $W_{31}$ that appears in individual Feynman integrals but (as far as is currently known) drops out of every suitably defined finite quantity in four dimensions; perhaps the same is true of the letters $W_{[123,128]}$ and $W_{138}$ at six points.

Therefore 36 rather mysterious letters remain, including $W_{[52,57] \cup [70,75] \cup [100,105] \cup [242,247]}$ which appear at $\mathcal{O}(\epsilon^0)$ in dimensional regularization, and more urgently $W_{[58,69]}$ which appear already at $\mathcal{O}(\epsilon^{-1})$. We defer the search for a satisfactory ``explanation'' of these letters for future work; 24 of these 36 letters appear in the maximal transcendental part of the six-point three-loop all-plus amplitude in pure Yang-Mills theory~\cite{Carrolo:2025pue}, but based on the very limited information available at the moment we cannot exclude the possibility that some or all of the remaining 12 ($W_{[52,57] \cup [70,75]}$) might drop out of suitably defined finite quantities in four dimensions.

There is one more interesting thing to note about these last 12 letters. In our paper we have used the $\Fl_{2,n-2;n}$ parameterization of massless kinematics in terms of spinor helicity variables. The recent paper~\cite{Bossinger:2025rhf} also considers the $\Fl_{2,4;n}$ parameterization using momentum twistors and an infinity twistor. For $n=6$ the two approaches happen to both involve $\Fl_{2,4;6}$, but they are non-isomorphic.  Nevertheless, the conclusions summarized in the above table are almost identical between the two approaches; the only exceptions are the 12 letters $W_{[52,57] \cup [70,75]}$ which are not cluster variables of $\Fl_{2,4;6}$ when the spinor helicity embedding is used but are (ratios of) cluster variables when the momentum twistor embedding is used~\cite{Bossinger:2025rhf}.

We also note that (the cyclic closure of) the five-point one-mass two-loop planar symbol alphabet~\cite{Abreu:2020jxa} is naturally a subset of the six-point massless two-loop alphabet, but even this subset includes letters that are not products of $\Fl_{2,4;6}$ cluster variables, specifically: $W_{[52,75] \cup [123,128]}$.

Our paper opens several additional directions for further work.
It would be interesting to explore cluster adjacency in the context of $\Fl_{2,4;6}$; cluster adjacency~\cite{Drummond:2017ssj} refers to the observed connection between the mathematical notion of cluster compatibility (which pairs cluster variables appear together in a common cluster) and which pairs of symbol letters can appear next to each other in a symbol. Cluster adjacency for the two-loop five-point planar alphabet was described in~\cite{Bossinger:2022eiy}.

It would be interesting to explore the connection between $\Fl_{2,4;6}$ cluster variables and the facets of tilings of the momentum amplituhedron~\cite{Damgaard:2019ztj}; for the momentum twistor amplituhedron that describes the amplitudes of tree-level SYM theory it has been proven~\cite{Even-Zohar:2024nvw} there is a deep connection between tilings of the $n$-particle amplituhedron and (compatible sets of) cluster variables of $\Gr(4,n)$.

Finally, it would be interesting to seek out more data on massless symbol alphabets, for example for the non-planar sector of two-loop six-point integrals, or at higher-loop order or higher number of external points; for example, recently the complete basis two-loop five-point two-mass (effectively a subset of the two-loop seven-point) planar integrals has been constructed in~\cite{Abreu:2024yit} where there are 286 algebraic letters and 284 rational letters. At higher-loop order, some massless three-loop five-point integrals have been computed in~\cite{Liu:2024ont} while 30 new planar letters are conjectured in~\cite{Chicherin:2024hes}.
In addition, in the non-planar integral section, the two-loop five-point one-mass family~\cite{Abreu:2021smk, Abreu:2023rco} and a subset of three-loop four-point one-mass integrals~\cite{Gehrmann:2024tds} were evaluated.

\acknowledgments

We are grateful to S.~Abreu, L.~Bossinger, L.~Dixon, J.~Drummond, R.~Glew, \"O.~G\"urdo\u{g}an, J.~Henn, J.-R.~Li, E.~Mazzucchelli, R.~Wright and Y.~Zhang for helpful discussions and correspondence. This work was supported in part by the US Department of Energy under contract DE-SC0010010 Task F and by Simons Investigator Award \#376208 (AV). Part of this research was conducted using computational resources and services at the Center for Computation and Visualization, Brown University.

\bibliographystyle{JHEP}

\bibliography{biblio.bib}

\end{document}